# One hundred years ago Alfred Landé unriddled the Anomalous Zeeman Effect and presaged Electron Spin

Horst Schmidt-Böcking[a] and Bretislav Friedrich[b]

(a) Institut für Kernphysik, Universität Frankfurt, Max-von-Laue-Str.1, D-60438 Frankfurt, Germany
(b) Fritz-Haber-Institut der Max-Planck-Gesellschaft, Faradayweg 4-6, D-14195 Berlin, Germany

*Dedicated to Steven T. Manson on the occasion of his 80th birthday*

**Abstract:**

In order to commemorate Alfred Landé's unriddling of the anomalous Zeeman Effect a century ago, we reconstruct his seminal contribution to atomic physics in light of the atomic models available at the time. Landé recognized that the coupling of quantized electronic angular momenta via their vector addition within an atom was the origin of all the apparent mysteries of atomic structure as manifested by the anomalous Zeeman effect. We show to which extent Landé's ideas influenced the development of quantum physics, particularly Wolfgang Pauli's path to the exclusion principle. We conclude with Landé's brief biography.

|  **Der Drall*** | **The Swirl**** |
|---|---|
| *Mit Drall hat Gott die Welt erschaffen* | *With a swirl did He make all spin* |
| *Vom Sternenmeer zum Elektron* | *From the sea of stars to the electron* |
| *und ärgert auch der Tanz den Pfaffen,* | *Whose dance makes the vicar think of sin,* |
| *Ei, Gott dem Herrn gefällt er schon.* | *But pleases Him, The Sine-Qua-Non.* |
| | |
| *Drum dreh dich, dreh, mein wildes Mädel,* | *So twirl, twirl, my fierce girl,* |
| *Geschwind, bis uns der Atem fehlt.* | *Quick, before we run out of breath.* |
| *So schnell wie wir dreht sich kein Rädel,* | *No wheel turns as fast as we do,* |
| *Wenn Tanzmusik uns ganz beseelt.* | *When the band's blast goes whop-de-do.* |
| | |
| *So schnell wie wir dreht sich kein Wirbel,* | *As fast as we turns no whirl,* |
| *Von Wind und Welle angefacht,* | *Fanned by wind and wave,* |
| *Wenn von der Zehe bis zur Zirbel* | *When from head to toe* |
| *Ein jeder Muskel tanzend lacht.* | *All muscles jig and swerve.* |
| | |
| *So schnell wie wir walzt keine Spindel* | *No spindle spins as fast as we have* |
| *Und wickelt ihren Faden ab.* | *And unwinds its curly fiber.* |
| *Die Mutter dreht uns in der Windel,* | *Mother winds us in the diaper,* |
| *Die Erde dreht uns noch im Grab.* | *And Earth keeps us turning in the grave.* |

*This poem was written by the 1966 Nobel Laureate **Alfred Kastler** (1902-1984), on the occasion of the 80th birthday of Adalbert Rubinowicz (1889-1974). on 22 Februar 1969. It was published in Ref.[1] "Drall" is an obsolete German word for "angular momentum."
Figure 1 illustrates the awe angular momentum inspires.

** Translated by Bretislav Friedrich

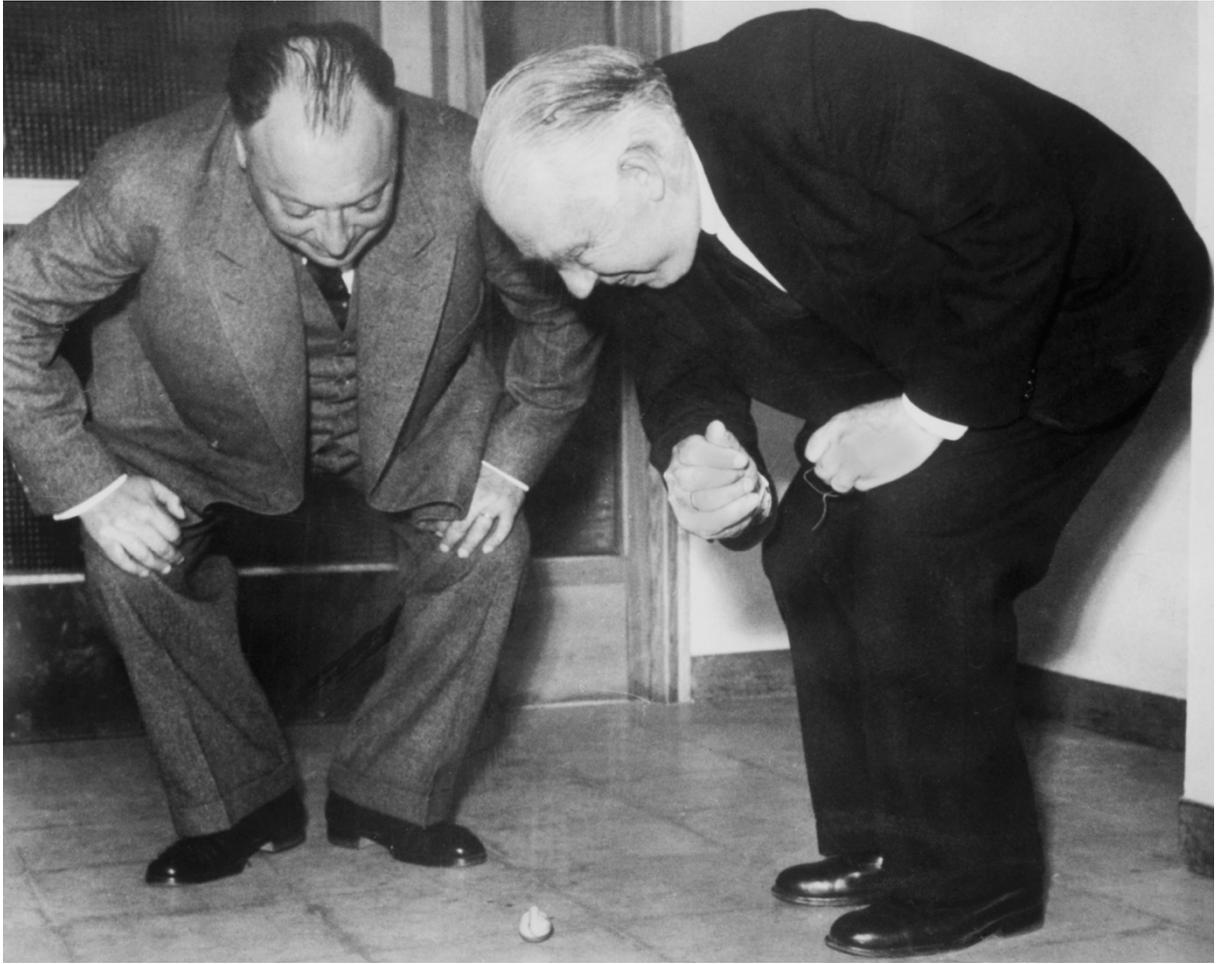

Figure 1: Captivated by the dynamics of the *Drall* (angular momentum), Wolfgang Pauli and Niels Bohr watch the spinning of a tippe-top during a break at the 1954 inauguration of the Institute of Physics, Lund, Sweden. Courtesy of Niels Bohr Archive (Copenhagen), photographic collection.

# 1. Introduction

A century ago, the anomalous Zeeman effect (AZE)[2,34,5] manifest in the multiplet structure of atomic spectra was a big riddle. Physicists such as Niels Bohr (1885-1962), Arnold Sommerfeld (1868-1951), Pieter Debye (1884-1966), Hendrik Kramers (1894-1952), Werner Heisenberg (1901-1976), Wolfgang Pauli (1900-1958), Alfred Landé (1888-1976) and others did their utmost to unravel the "number mystery" of the AZE, as Sommerfeld dubbed the riddle.[6] Pauli would chime in later and call it the "big misery [of the AZE]," Ref.[7] p. 203.

It seems that the terms *normal* and *anomalous* Zeeman effect were coined by Friedrich Paschen and Ernst Back in Ref.[3]. The terms normal and anomalous Zeeman effect were adopted – and "codified" – by Arnold Sommerfeld in Ref.[4].

Despite "tailored" postulates contradicting classical physics, such as electron orbits in atoms with quantized angular momentum (and energy) introduced by Bohr[8] or space (directional) quantization of the planes of the electron orbits and thereby also of the atomic magnetic moments in an external magnetic field introduced by Sommerfeld[9] and Debye[10], all attempts up to 1921 to unriddle the anomalous Zeeman effect failed.

It was Alfred Landé, Figure 2, who succeeded in Frankfurt in 1921 to unravel the anomalous Zeeman effect, relying on empirical methods[11,12] rather than on first principles. In order to explain the regularities gleaned from the number mystery (multiplet structure and frequency values of the spectral lines), Landé introduced additional postulates that contradicted classical physics. Not surprisingly, these postulates, which Landé referred to as *Arbeitshypothesen* [working hypotheses], were not readily accepted by the "patriarchs" of atomic physics at the time, such as Sommerfeld or Bohr. The postulates were:[11,12]

1. In multi-electron atoms, such as helium, the angular momenta of the electrons combine vectorially and their projections on an external magnetic field obey the space-quantization rules.
2. Atomic angular momentum quantum numbers can take half-integer values.
3. The gyromagnetic behavior of electronic angular momenta does not obey classical electrodynamics. Deviations from the laws of classical physics are accounted for by the g-factor.

For the singlet states of atoms (i.e., for states with zero electron spin), the dimensionless gyromagnetic ratio (the ratio of the magnetic moment expressed in Bohr magnetons to the angular momentum) – the g-factor – is equal to *one*, but for higher-multiplicity states (i.e., states with non-zero electron spin) it can take on different values. In his empirical quest to understand the AZE that predated the 1925 discovery of electron spin[13], Landé was able to find an accurate expression for the g-factor in terms of a ratio of integer or half-integer quantum numbers. For the half-integer angular momentum ½ ℏ, Landé's expression yielded the g-factor of *two*. Landé's results were recently extolled by Friedrich et al.[14] as important milestones on the path to unriddling the AZE as well as to discovering electron spin and the Pauli principle.

It seems that Landé's results had not been as widely recognized as pioneering achievements as they would have deserved. Was it the *ad hoc* character of his empiricism that was turning off his colleagues in theoretical physics? Even Landé's mentor, Max Born (1882-1970), with whom he shared an office in Frankfurt, scoffed at his work on the AZE. In his 1962 interview with Thomas Kuhn, Born described the situation as follows:[15]

> *Then he [Landé] came to Frankfurt again, and his head was completely occupied with the paper which I didn't grasp at first. It was these whole number relations between the intensities of multiplet lines and Zeeman-effect lines. And he did it in a way which seemed to me horrible, namely, simply by guessing about numerical values. ... At last came a formula which gave all the results he wanted. I couldn't check it -- I can never do numerical calculation problems. So I didn't take much notice of him, and he also did not take much notice of our work, though we were sitting all the time in the same room. But two years later, or three, when we derived the square root of integers formula from quantum mechanics, we saw at once that it was very important.*

Landé went on to correspond with Sommerfeld and Bohr, but neither agreed with his ideas, in particular those concerning half-integer quantum numbers. On the other hand, Heisenberg[16] and Pauli[17] supported Landé's arguments and Heisenberg even wrote a joint paper[18] with Landé on the AZE. For Pauli, Landé's results were pace-setting achievements. In his Nobel Lecture, Pauli characterized Landé's pioneering work as "of decisive importance for the finding of the exclusion principle."[19]

When the matrix and wave mechanics of Heisenberg and Erwin Schrödinger (1887-1961) proved consistent with Landé's g-factor formula and his working hypotheses, the new theories (of spin-orbit coupling and of angular momenta) only confirmed what Landé had conjectured for years.

Perhaps with the benefit of hindsight, Friedrich Hund's 1975 "History of Quantum Theory"[20] – a first-hand account in many respects – detailed Landé's contribution and its ramifications for the discovery of electron spin.

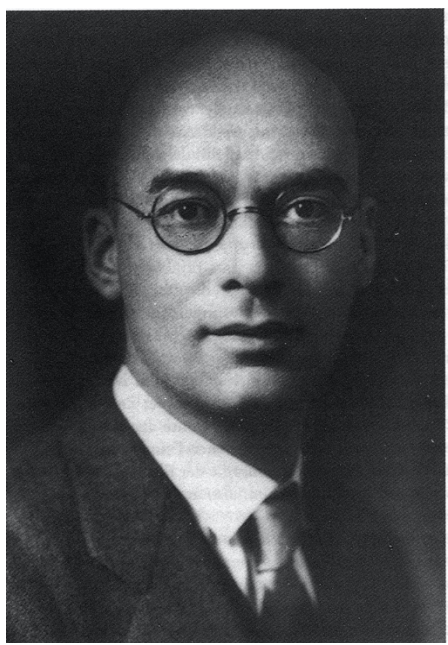

Figure 2: Alfred Landé 1920 in Frankfurt. Courtesy of Universitätsarchiv Frankfurt am Main.

Paul Forman (*1937) in Refs.[17,21] as well as Jagdish Mehra (1931-2008) together with Helmut Rechenberg (1937-2016) in Ref.[7] described in great detail the numerous attempts by Sommerfeld, Bohr, Heisenberg, Pauli, Landé and others to make sense of the AZE. One of the main riddles was: where in the electronic shell of many-electron atoms does the atomic magnetic moment come from – from the outer electron or from *Rumpf*, the inner core? Landé and others thought in 1921 it was the core. Moreover, it was unknown which electrons carry which angular momentum, and how many electrons can fit in the same shell. Thus, there was quite some arbitrariness and confusion in assigning quantum numbers and angular momenta to electrons in atoms. That the notation of angular momenta varied from author to author and from paper to paper further exacerbated the situation.

In contrast to the aforementioned accounts[7,17,21] our goal is not to survey past attempts to understand the Old Quantum Theory of the atom. Instead, we provide today's perspective on Landé's important findings and thereby make Landé's empirical path to unraveling the number mystery of the AZE easier to follow.

Asim Orhan Barut (1926-1994), a close coworker of Landé, extolled Landé's pioneering contributions:[22]

> *The Frankfurt period [1919-1922] ... of ALFRED LANDÉ was perhaps the most important period of his working life as a scientist. He became particularly well-known through the works from this time, which secured his enduring fame. ... The most burning problem in physics at the time was the task of explaining the anomalous ZEEMAN effect. SOMMERFELD had just devoted a long paper to it, upon which many young physicists turned to tackling the problem. After a visit to BOHR in Copenhagen in October 1920, LANDÉ too began, in December, to seriously study the problem of the anomalous ZEEMAN effect.*

## 2. Models of atomic structure a century ago and their forerunners

What was known before 1921 about the internal structure of atoms? The measured quantities of atomic structure available at that time were the complex multiplet structures of the spectral lines and their corresponding wavelengths, their splitting and shifts in electromagnetic fields (Zeeman[2], Stark[23], and Paschen-Back[3] effects) and their polarization. Figure 3 presents the timeline of the milestones on the path to understanding atomic structure. Among the experimental ones were the spectral analysis by Robert Wilhelm Bunsen (1811-1899) and Gustav Robert Kirchhoff (1824-1887), who showed that atoms absorb and emit light of characteristic wavelengths.

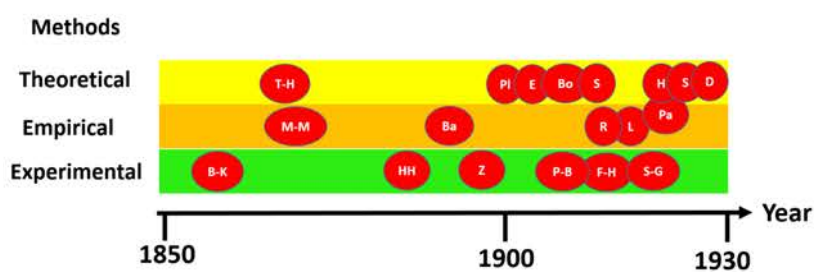

Figure 3: Timeline of milestone discoveries – theoretical, empirical, and experimental – in the atomic structure research. B-K stands for Bunsen and Kirchhoff; H-H for Hertz and Hallwachs; Z for Zeeman; P-B for Paschen and Back; F-H for Franck and Hertz; S-G: for Stern and Gerlach; M-M for Mendeleev and Meyer; Ba for Balmer; R for Rubinowicz; L for Landé; P for Pauli; T-H for Thomson and Helmholtz; Pl for Planck; E for Einstein; Bo for Bohr; S for Sommerfeld; H for Heisenberg; S for Schrödinger, and D for Dirac (see text).

In 1886, Heinrich Hertz (1857-1894) and Wilhelm Hallwachs (1859-1922) found that electrons can be released from surfaces by light whose wavelength is short enough[24,25]. Pieter Zeeman (1865-1943) discovered in 1896 that spectral lines split up in an external magnetic field and are shifted in accordance with the field's strength[2]. Moreover, James Franck (1882-1964) and Gustav Hertz (1887-1975) were able to show in 1914 that the energy levels of electrons bound in atoms are discrete[26].

In 1922, the Stern-Gerlach experiment[27] provided convincing evidence corroborating several of the proposed postulates related to the electronic angular momentum and magnetic moment of atoms. Details will be discussed below.

The development of atomic models was also advanced by chemists, especially by Lothar Meyer (1830-1895)[28] and Dmitri Mendeleev (1834-1907)[29] whose periodic table of the elements presaged the discovery of closed shells. In 1986, Johann Jakob Balmer (1825-1898) empirically found an accurate formula for one of the spectral series of hydrogen[30]. Balmer's work paved the way for Bohr's model of the atom. Systematic spectroscopic investigations enabled the determination of energy levels of other atoms. In 1918, Wojciech Rubinowicz (1889-1974) discovered that optical transitions could only take place if certain conditions were met by the quantum numbers of the initial and final states involved and introduced selection rules $\Delta m = m_e - m_g = 0, \pm 1$ for the difference between the magnetic quantum numbers $m_e$ and $m_g$ of the excited and ground states[31]. Why such selection rules exist became clear when it was recognized that every photon has an angular momentum of $h/2\pi \equiv \hbar$ and therefore only certain states can be "combined" with one another because of conservation of angular momentum. Rubinowicz's selection rule for allowed Zeeman transitions was crucial for Landé's postulation of half-integer quantum numbers. In 1926-27, Wigner was the first to employ group-theoretical considerations for interpreting the selection rules of atomic spectroscopy. He accomplished this by invoking the transformation properties of energy eigenstates of a system with respect to operations which leave the system unchanged (space rotations, mirror inversions, permutations of the electrons). The encounter between group theory and the old-quantum-theoretical notion of "selection rules" had a profound and long-lasting impact on the physical content of quantum theory.[32]

Nearly all early atomic models described below assumed particulate constitution of matter – with the constituents either static or in motion. However, one of the earliest scientific models of the atom, the vortex atom, was inspired by the mechanics of continua as developed by Hermann von Helmholtz (1821-1894). In 1867, William Thomson (Lord Kelvin) (1824-1907) made use of Helmholtz's vortices and vortex filaments of a hypothetical fluid to conceive a model of matter that represented different types of atoms by different knots formed by the vortex filaments[33]. The knot theory was advanced by Kelvin's friend and the translator of Helmholtz's treatise[34] into English, Peter Guthrie Tait (1831-1901)[35]. Given that the prime candidate for the fluid whose dynamics gave rise to the knotted filaments was the luminiferous ether, the vortex model was intimately connected with the electromagnetic theory of James Clerk Maxwell (1831-1879). This connection provided an impetus for the search – and eventual discovery – of the electron by Joseph John Thomson (1856-1940) in 1897. The vortex theory was abandoned by both Kelvin and J.J. Thomson before the end of the 19th century because of its inconsistencies and inability to account for gravity, among other reasons. As Helge Kragh noted, "Many theoretical physicists believe today that some version of superstring theory may accomplish what the vortex theory could not in the past – and much more"[36].

The discovery of the quantum of action, h, by Max Planck (1858-1947) in 1900[37] as well as its "second coming" (Abraham Pais' term[38]) in the quantum theory of the electromagnetic field by Albert Einstein (1879-1955)[39] in 1905 paved the way for "the third coming" of h in Bohr's model of the atom[40]. By postulating electron orbits with discrete angular momenta $|\mathbf{k}|=k\hbar$, with k=1,2,3, … the azimuthal quantum number, and by invoking the combination principle of Johannes Rydberg (1854-1919)[41] and Walter Ritz (1878-1909)[42], Bohr was able to derive the Balmer series as well as to interpret the Rydberg constant in terms of fundamental constants, including h. From then on, the quantized angular momentum of the electron(s) has been the mainstay of atomic structure.

In Bohr's model, an electron with angular momentum $\mathbf{k}$ revolving on a circular orbit about the heavy nucleus generates a magnetic moment $\boldsymbol{\mu} = -e/(2m_ec)\mathbf{k}$ such that $\mu \equiv |\boldsymbol{\mu}| = k\mu_B$, where $\mu_B \equiv e\hbar/(2m_ec)$ is the Bohr magneton, with e the magnitude of electron's charge, $m_e$ the electron mass, and c the speed of light. According to the laws of classical physics, in an external magnetic field $\mathbf{B}$, this atomic magnetic moment would perform Larmor precession of frequency $\nu_0 \equiv 1/(2\pi)eB/(2m_ec)$, with the angle α spanned by the vectors $\mathbf{B}$ and $\boldsymbol{\mu}$ remaining constant. The Zeeman energy of the orbiting electron due to its interaction with the magnetic field would then be $W = -k\mu_B B \cos\alpha$, where $B \equiv |\mathbf{B}|$. For equiprobable directions α, the Zeeman energy of an atomic state was therefore expected to broaden. Same for transitions between such states. However, the experimental observations showed that the spectral lines remained sharp and that they split into multiplets (singlets, doublets, triplets, etc.). The singlets and triplets were referred to as due to a normal Zeeman effect (NZE), the rest as due to an anomalous Zeeman effect (AZE).

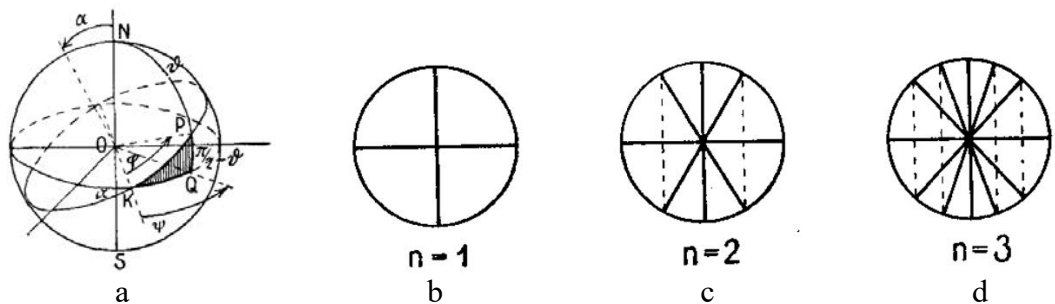

Figure 4: (a) Spherical coordinate system and an electron trajectory on the unit sphere passing through points K and P as envisioned by Sommerfeld; (b)-(d) Allowed projections of the orbital angular momentum for a given azimuthal quantum number denoted here as n on a horizontal magnetic field. Adapted from Sommerfeld A "Zur Quantentheorie der Spektrallinien" 1916 Annalen der Physik 51, 1-94, Ref.[43].

Sommerfeld[9] and Debye[10] proposed as explanation for the sharp lines in the Zeeman spectra the space quantization (*Richtungsquantelung*) of the electronic angular momenta, whereby only discrete directions α of $\mathbf{k}$ with respect to $\mathbf{B}$ would be allowed. The electron orbits, assumed to be perpendicular to the angular momenta, would have to assume discrete spatial orientations as well. Figure 4 illustrates the possible orbital directions in a horizontal magnetic field (panels b-d) for different n. The projection of the discrete electronic orbital angular momentum on the magnetic field is then given by the magnetic quantum number $m = n\cos\alpha$ and hence the energy due to the normal Zeeman effect by

$$W = -k\mu_B B \cos\alpha = h\nu_0 m \qquad (1)$$

In 1920, from his analysis of the AZE, Sommerfeld concluded[4] that an additional, fourth quantum number, j, must exist, which he called the inner quantum number. In Ref.[4], page 231, Sommerfeld wrote:

*The distinguishing feature of the various d [k=2] and p [k=1] terms must rather be an* inner *quantum number [our emphasis], corresponding perhaps to a hidden rotation.*

On page 234 he continued:

*I would like to emphasize that we proceeded quite formally in applying our selection principle to the "inner" quantum numbers as well as their very choice. The physical fact is that certain term combinations do not arise.*

On page 238, Sommerfeld discussed half-integer quantum numbers:

*For the quantum theory of spectral lines, which works with integers, the running number m +½ of the s-term is a problem… It was only the connection of the selection principle with the term combinations that prompted me to assign the value 1 to the quantum number of the s-term, but at the same time to doubt whether the notation m +½ is really appropriate.*

But then he found reasons to exclude half-integer quantum numbers based on experimental data. Nevertheless, the considerations by Sommerfeld provided hints and stimulating ideas for Landé to formulate his *"Arbeitshypothesen"* (working hypotheses) that would lead him to his take on half-integer quantum numbers and the g-factor.

In 1916 Sommerfeld further refined Bohr's model of the atom. Expanding on the legacy of Karl Schwarzschild (1873-1916)[44], he succeeded in explaining the fine structure of the spectral lines. To this end, Sommerfeld introduced elliptical orbits of the electrons, whose eccentricity he characterized with another quantum number. By taking into account the relativistic mass increase of the electrons on an elliptical orbit in the perihelion phase, Sommerfeld was able to accurately describe the fine-structure splitting of spectral lines.

Only after 1925 when it became clear that the electron itself has a magnetic moment with an angular momentum of ½ ℏ and a g-factor of about 2 and that the angular momenta couple due to magnetic interaction to a quantized total angular momentum *J,* the number mystery and misery of the AZE could finally be put to an end by making use of the Pauli principle and the definitive quantum theories of Heisenberg and Schrödinger.

## 3. Landé's atomic structure models and vector coupling of atomic angular momenta

Landé's interest in the AZE was likely triggered by his discussions with Bohr during his visit to Copenhagen in October-November 1920, Refs.[7,17,21]. Before then, Landé made forays into atomic structure, see items 9-27 in the "Selected Scientific Papers of Alfred Landé," Ref.[45]. In 1918, in four joint publications with Born,[46,47,48,49] he examined the question of whether atoms are planar (2D) – as suggested by the Bohr model – or three-dimensional (3D). Based on classical calculations of static charge distributions, Born and Landé concluded that only 3D atoms could give rise to crystal structures with the observed stability and compressibility. Building in part on chemical heuristics, especially that gleaned from the chemistry of carbon, Landé continued this line of research in 1919 by examining the stability of the underlying multielectron atoms. Assumed to be in motion but at fixed separation from each other, Landé's electrons congregated in various polyhedral,[50,51,52] including the tetrahedron and the cube[53,54,55,56]. In Ref.[57], Landé examined the dynamics of electrons obeying Bohr's quantization conditions and found collective orbits, Figure 5, that were stable. The question remained as to *how* stable these orbits were with respect to initial conditions and perturbations, such as electron capture. The angular momentum balance was not yet considered in these works.

Also in 1919, Landé examined for the first time the electronic angular momenta and the spatial orientation of two elliptical electron orbits in the helium atom[58]. In a follow-up work[59] that went into his 1920 *"Habilitationsschrift,"* Landé calculated the electron orbits and the electronic energy of the He atom using perturbation theory.

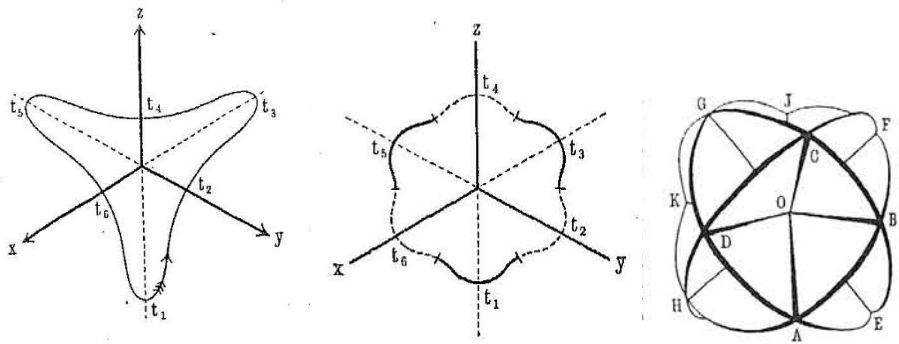

Figure 5: Examples of atomic electron trajectories as proposed by Landé A in "Dynamik der räumlichen Atomstruktur" 1919 Verhandlungen der Deutschen Physikalischen Gesellschaft 21 2-12; 21 644-652; 21 653-662, Ref.[57].

By careful examination of the available spectroscopic data, Landé was able to infer that the angular momenta of the core (inner) electron and of the luminous (outer) electron combine via vector addition, $\mathcal{J} = J + J'$ (see Figure 6). Landé presumed that the total angular momentum, $\mathcal{J}$, of the atom in an external magnetic field was subject to the quantization conditions, including space-quantization.

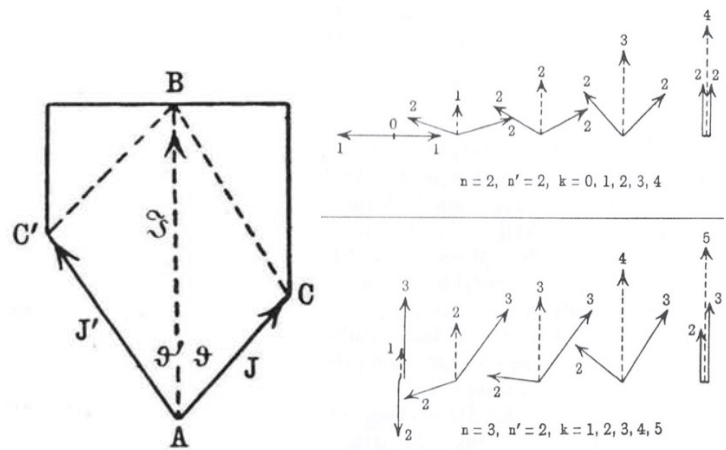

Figure 6. Left: Vector addition of the electronic angular momenta J' and J (pertaining, respectively, to the inner and outer electrons) to the total angular momentum ***J***. The projections of the angular momenta in the direction of the magnetic field are integer multiples of h/2π. Right: Vector addition of orbital angular momenta J' and J (shown by arrows) of lengths given by quantum numbers n and n' that add up to a total angular momentum k (dashed arrow). Adapted from Landé A "Eine Quantenregel fur die räumliche Orientierung von Elektronenringen" 1919 Verhandlungen der Deutschen Physikalischen Gesellschaft 21, 585-588, Ref.[58].

## 4. Landé's working hypothesis about the existence of half-integer quantum numbers

In 1921, Landé made a breakthrough in explaining the observed splitting of spectral lines due the AZE which he reported in these two papers: *On the anomalous Zeeman effect (Part I)*[11] *(submitted on April 16, 1921)* and *On the anomalous Zeeman effect (Part II) (submitted on October 5, 1921)*[12]. In order to describe the number of lines observed for doublets and quartets, Landé identified Sommerfeld's "inner" quantum number[4] of an atom with the quantum number j of the atom's total angular momentum and allowed the projection quantum number m to take half-integer values.

Why half-integer values? Building on experimental observations, Landé assumed that the selection rule $\Delta m = 0, \pm 1$ for spectral lines polarized, respectively, parallel and perpendicular to the Zeeman field has to apply to the NZE and the AZE alike and that the splitting of the ± m lines in a magnetic field must be symmetric with respect to the line position at zero magnetic field. Landé wrote in Ref.[11], p. 234:

> *While the usual space quantization in a magnetic field admits only integer values of m, one must come to grips here with rational fractions of m* [justified in Ref.[12] by "anomalous" Larmor precession] *such ... that adjacent values of m are separated by ± 1. Because of the + and − symmetry, the only possible sequence of fractions is: m = ±1/2, ±3/2, ..., ±(2j−1)/2, apart from the other* [integer] *sequence m = 0, 1, 2,..., j* [where we, for consistency, wrote j instead of Landé's J].

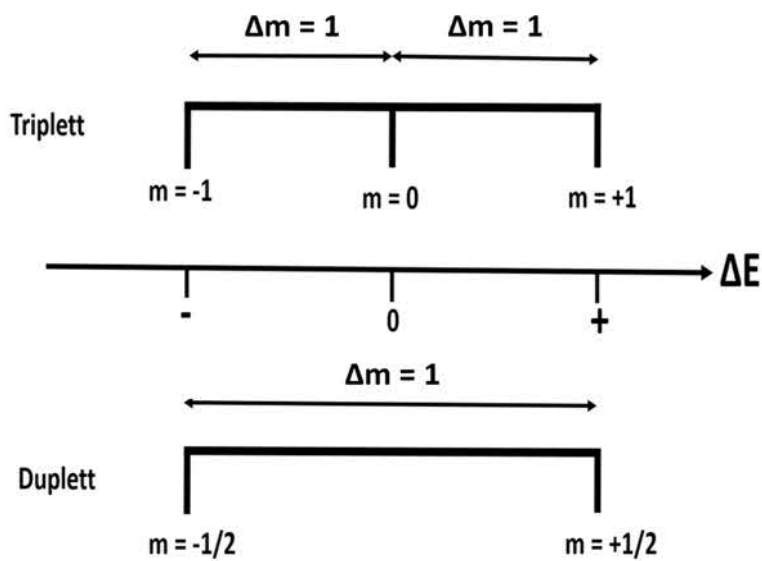

Figure 7. Top: Energy splitting in the case of the normal Zeeman effect (Triplet). Bottom: Energy splitting in the case of the anomalous Zeeman effect (a doublet).

Landé's argument is illustrated in Figure 7, whose upper part shows the triplet splitting due to the NZE. In accordance with the selection rule for Δm, three lines are observed: one at the origin (m = 0), with zero energy shift, flanked by two symmetrically shifted lines corresponding to m = +1 (right) and m = −1 (left). The lower part of Figure 7 exemplifies the AZE for the case of a doublet: only two lines are observed with no line at the origin (given by the line position at zero magnetic field). In order to comply with the Δm = ± 1 selection rule, the two symmetric lines then must have half-integer quantum numbers, namely m = + ½ or m = − ½.

## 5. Landé's g-Factor

On his path to a *common* formula for the normal and anomalous Zeeman effects, Landé relied on the following types of spectroscopic data measured for atoms subject to a magnetic field: (a) the number of lines in a multiplet series in the magnetic field; (b) the relative line intensities; (c) the absolute energy shifts of the lines in the magnetic field; (d) the polarizations of the lines. In addition, Landé made use of the Rydberg-Ritz combination rule and of the selection rule of Rubinowicz for Δm.

During the period 1920-1924, when Landé tackled the AZE, he had not entertained the idea that electrons may carry their own, internal, angular momentum (spin) and a magnetic moment associated with it. And neither had he considered the possibility that the half-integer quantum numbers he found may pertain to the inner (spin) angular momentum of the electron. Instead, Landé presumed that the total magnetic moment of an atom is made up of the magnetic moment due to the orbital motion of the electrons in the *core* and of the magnetic moment due to the orbiting *outer* electron(s). This viewpoint is summarized in a 213-page review *The Zeeman effect and the multiplet structure of spectral lines*[60] written jointly in autumn 1924 with his erstwhile competitor Ernst Back (1881-1959).

In order to treat the anomalous Zeeman effect, which would encompass all the angular momenta and associated magnetic moments involved, Landé made the following *Ansatz*: he divided the Zeeman energy of Eq. (1) by the Larmor energy, $h\nu_0$, obtaining $W/(h\nu_0) = m$. Then he introduced a factor g,

$$W/(h\nu_0) = mg \qquad (2)$$

that led to the normal Zeeman effect for g = 1 and rendered any deviations from it – i.e., the anomalous Zeeman effect – for any other value of g. Eq. (2) then indeed amounts to a common formula for the NZE and AZE. Landé referred to the g-factor as the "splitting factor." As we know today, Landé's g-factor encompasses the entire electronic angular momentum structure – orbital and spin – of an atom.

For the NZE, the values of m are m = –1, 0, +1 for j = 1 and m = –2, –1, 0, +1, +2 for j = 2, resulting in g = 1. However, for the AZE, i.e., half-integer quantum numbers (equation (7) of Ref.[11]), m = ± 1/2; ± 3/2; ... ± (2j–1)/2. Thus, it followed that by setting mg = k,

$$g = 2k/(2j-1) \qquad (3)$$

for m = (2j–1)/2. For the doublets (with k = j = 1), the g-factor formula (3) then yielded the g-factor

$$g = 2 \qquad (4)$$

see equation (11) of Ref.[11]. Although apparently arrived at empirically, Landé gave a rationalization for equating mg with k in his subsequent paper[12].

Sommerfeld had concluded in Ref.[4] that the term denominator for doublets should be 2j-1 as in Landé's formula, Eq. (3). But he ruled out half-integer quantum numbers and it was on that occasion that he called the treatments of the AZE a "number mystery." Further examples of g-factors calculated from equation (3) are given in Table 1.

| Term | s | p1 | p2 | d1 | d2 |
|------|---|----|----|----|----|
| k    | 1 | 2  | 1  | 3  | 2  |
| j    | 1 | 2  | 2  | 3  | 3  |
| g    | 2/1 | 4/3 | 2/3 | 6/5 | 4/5 |

Table 1: Examples of g-factors calculated from Landé's formula, Eq. (3) of Ref.[11].

Landé concluded his first paper on the AZE[11] by noting that making sense of the AZE is tantamount to understanding the origins of the g-factor. In the sequel[12], submitted five months later, he wrote:

*After the complicated Zeeman types were reduced to the anomalous basic factors g not equal to 1 in Part I, an attempt will now be made to find out whether the anomaly of g can be traced to a modification of Larmor's theorem for the precession of electron systems in a magnetic field.*

Referring to the spectra of barium[61] obtained by Frederick Albert Saunders (1875-1963) and of caesium[62] reported upon by Karl Wilhelm Meissner (1891-1959) as well as to the experiments by Jackson Barnett (1873-1956)[63] and Einstein-de Haas (Wander Johannes de Haas (1878-1960))[64], Landé came to the conclusion that there must be a previously unknown additional rotation of the atom that "turns on" in a magnetic field. He likened this rotation to that of the Foucault pendulum in the earth's rotating gravitational field and attributed a half-integer angular momentum and g-factor of 2 to it. Landé thus gathered most of the ingredients of electron spin – except for recognizing that the half-integer angular momentum and the anomalous g-factor had to be ascribed to the electron itself.

Landé continued analyzing ever more accurate (about 5%) and diverse spectroscopic data that included the multiplets (up to octets) discovered in 1922 by Miguel Catalan (1894-1957)[65] and Hilde Gieseler[66]. In his empirical search for additional regularities in the dependence of the g-factor on the spectroscopic observables, Landé eventually redefined the set of the quantum numbers in terms of which the g-factor was expressed. In 1923, he inferred the following relationship between the g-factor, the multiplicity r (number of lines in a multiplet, i.e., r = 1,2,3 for singlet, doublet, triplet …), the azimuthal quantum number k (s, p, d, …), and the maximum magnetic quantum number $|k–1– (r–1)/2| \leq m_{max} \leq |k–1+(r–1)/2|$ (see Ref.[5], page 157):

$$g = \frac{3}{2} + \frac{\left(\frac{r}{2}\right)^2 - \left(k - \frac{1}{2}\right)^2}{2(m_{max}+1)m_{max}} \qquad (5)$$

The consolidation of Landé's views on the AZE took a new turn when, in a duo of 1923 papers *Term structure and the Zeeman effect of multiplets*[67,68] he defined a new triad of quantum numbers – K, R, J – to characterize an atomic term. Apparently encouraged by the partial success of the g-factor of Eq. (5), Landé obtained this triad by mapping k–1/2 onto K, r/2 onto R, and j onto J for even multiplets and j–1/2 onto J for odd multiplets. The addition of the corresponding vectors **K** and **R** then resulted in the following range of the quantum number pertaining to the total angular momentum **J**:

$$J = K+R-1/2, K+R-3/2, \ldots, |K-R|+1/2 \qquad (6)$$

which, in turn, led to the values of the projection quantum number $m = J-1/2, J-3/2, \ldots, -J+1/2$ and term energies $W(R,K,J,m) = mg(R,K,J)h\nu_0$, with the g-factor

$$g = g(R, K, J) = 1 + \frac{J - \frac{1}{4} + R^2 - K^2}{2\left(J - \frac{1}{4}\right)} \qquad (7)$$

whose form, Eq. (8) of Ref.[67], Landé justified by the vector model of the atom and the replacement of the squares of the angular momentum quantum numbers with what he called their "geometric means:" $J^2 \mapsto (J-1/2)(J+1/2)$; $K^2 \mapsto (K-1/2)(K+1/2)$; $R^2 \mapsto (R-1/2)(R+1/2)$. Aspects of the vector model are expanded upon and exemplified in the remainder of Ref.[67] and detailed in Ref.[14].

According to Pauli[69], the g-factor of Eq. (7) was accurate within 1%. In their 1925 progress report[60] on tackling the mysteries of the Zeeman effect written on the eve of the discovery of electron spin, Landé and Back noted:

> *This formula [Eq. (7)] has arisen empirically through theoretical considerations and is fully confirmed by the measurements of Catalan [Ref.[65]] and Back [Ref.[70]] on the manganese spectrum and H. Gieseler [Ref.[66]] on the chromium spectrum.*

A key ingredient that led to the success of the g-factor of Eq. (7) was taking the contribution from the core twice, cf. Refs.[67,14].

In the second paper[68] of the 1923 duo, Landé treated the anomalous Zeeman effect in the strong field limit, i.e., when the Zeeman splitting exceeds the multiplet splitting as given by the coupling of the **R** and **K** vectors. The result were the term energies

$$W = h\nu_0 m_K + 2h\nu_0 m_R = h\nu_0 m_K g(K) + h\nu_0 m_R g(R) \qquad (8)$$

with $m_K$ and $m_R$ the projection quantum numbers of the azimuthal and core angular momenta such that $m = m_K + m_R$, see Eq. (9) of Ref.[68]. However, whereas for the first (azimuthal) term of Eq. (8), $g = g(K) = 1$, for the second (core) term, $g = g(R) = 2$, consistent with Landé's above assumption that led to the weak-field limit g-factor, $g(R,K,J)$, of Eq. (7). The conceptual difficulties connected with the unknown origin of the angular momentum of the atomic core and other issues thwarting the nascent quantum mechanics were aptly summarized and discussed in Landé's 1923 commentary[71].

As noted in Ref.[14], from the positions of the nonvanishing g-factors in Table I of Ref.[67], Landé came close to recognizing that $R-1/2 \mapsto S$ (in current notation) is of special significance: by letting, in addition, $K-1/2 \mapsto L$ and $J-1/2 \mapsto J$ (in current notation), he would have reproduced the table for the quantum-mechanically admissible values $|S-L| \leq J \leq S+L$. Indeed, substituting from the above for R, K, and J into Eq. (7) gives the g-factor in the form

$$\begin{aligned} g = g(R, K, J) &= 1 + \frac{J(J+1) + S(S+1) - L(L+1)}{2J(J+1)} \\ &= \frac{J(J+1) - S(S+1) + L(L+1)}{2J(J+1)} + \frac{2[J(J+1) + S(S+1) - L(L+1)]}{2J(J+1)} \end{aligned}$$

$$(9)$$

with S, L, and J the quantum numbers of the spin, **S**, orbital, **L**, and total, **J**, angular momenta whose lengths are $|\mathbf{S}| = [S(S+1)]^{1/2}\hbar$, $|\mathbf{L}| = [L(L+1)]^{1/2}\hbar$, and $|\mathbf{J}| = [J(J+1)]^{1/2}\hbar$. Replacing the factor of 2 in the numerator of the second term by the gyromagnetic ratio of the electron, $g_S \approx 2.0023$, Ref.[72], would then give the current g-factor.

A graphical representation of how the g-factor comes about is given in Figure 8 for the atomic $^2P_{1/2}$ state, i.e., a state with S = 1/2, L = 1, and J = 1/2. The lengths of the blue arrows correspond to the lengths $|\mathbf{S}| = 3^{1/2}/2$, $|\mathbf{L}| = 2^{1/2}$, and $|\mathbf{J}| = 3^{1/2}/2$ of the angular momenta **S**, **L**, and **J** whose vector sum forms (in this case) an equilateral triangle. The red arrows represent the set of the corresponding magnetic dipole moments, each of which has the opposite orientation with respect to the angular momentum that gives rise to it. For a choice of lengths such that $|\mathbf{L}| = |\boldsymbol{\mu}_L|$, the length of $\boldsymbol{\mu}_S$ is then twice the length of **S**, i.e., $|\boldsymbol{\mu}_S| = 2|\mathbf{S}|$. The total magnetic dipole moment, $\boldsymbol{\mu} = \boldsymbol{\mu}_L + \boldsymbol{\mu}_S$, precesses about the total angular momentum **J** with a constant projection $\boldsymbol{\mu}_J$ on it. By setting $\mu_B = 1$, the g-factor is given by the ratio of the lengths of $\boldsymbol{\mu}_J$ and **J**, i.e., $g = |\boldsymbol{\mu}_J|/|\mathbf{J}| = 2/3$.

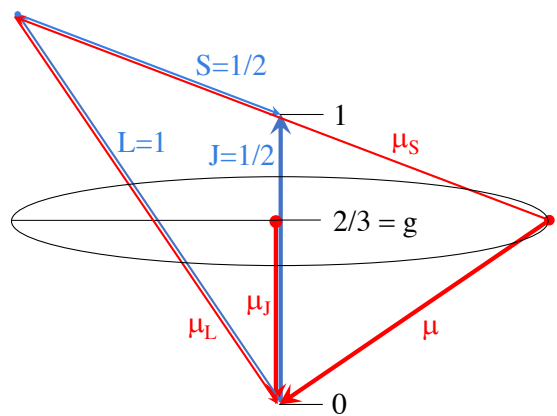

Figure 8: Graphical representation of the orbital, spin, and total angular momenta (blue) along with the associated magnetic moments (red) of an atomic $^2P_{1/2}$ state. Also shown is the resulting g-factor. See text.

## 6. Landé's 1923 interpretation of the Stern-Gerlach experiment

In 1923, within a year of the conclusion of the Stern-Gerlach experiment (SGE)[27,73,74,75,76,77], Landé made a connection of the SGE's outcome with the anomalous Zeeman effect[71]. Despite his daily contact with Otto Stern (1888-1969) and Walther Gerlach (1889-1979) during the period 1920-1922 when they toiled on the SGE at Frankfurt's Institute for Theoretical Physics, Landé's prescient interpretation of the SGE was barely noticed by anyone within the Institute or without.

The SGE was designed to see whether space quantization of atomic angular momentum was real and thereby "decide unequivocally between the quantum and classical views"[78]. The atomic angular momentum whose space-quantization was to be tested was the orbital angular momentum of the ground-state silver atom which was presumed to have k = 1. Some of the schools of thought leaned towards possible values of $m_K$=+1, 0, and −1 (represented by Sommerfeld) and some towards $m_K$=+1 and −1 (represented by Bohr). I any case, the silver atoms that would be deflected in an inhomogeneous magnetic field were expected to carry a magnetic dipole moment of 1 $\mu_B$. And this was indeed what the measured deflection appeared to confirm. However, what was at play was a "uncanny conspiracy of nature"[75]: the ground state of a silver atom is a $^2$S state that has zero orbital angular momentum (L=0) but a half-integer electron spin (S=1/2). Therefore, it was a combination of the corresponding half-integer projection quantum numbers ($m_S = \pm 1/2$) with the anomalous gyromagnetic ratio of the electron ($g_S \approx 2.0023$) that created the appearance of a magnetic moment $|\mu| = g_S \mu_B |m_S|$ of about 1 $\mu_B$ (experimentally, within 10%, Ref.[27]).

However, unlike Bohr, Sommerfeld, and pretty much everybody else, Landé, with his theory of the AZE, would not be fooled: he noted that for k=1, the silver beam would be split into three beams, corresponding to $m_K = -1$, 0, and +1. However, since a splitting into only two beams was observed in the SGE, Landé concluded that the silver atoms must be in a doublet state, with $m_R = -1/2$ and +1/2. The magnitude of the deflection would then correspond to 1 $\mu_B$ on account of the anomalous gyromagnetic ratio g = 2, $|\mu| = 2\mu_B |m_R| = \mu_B$.

Strangely enough, it was as late as 1927 when Ronald Fraser (1899-1985) determined that the ground-state orbital angular momentum and the associated magnetic moments of silver, hydrogen, and sodium are zero[79] and thus the doublet splitting seen in the SGE had to be attributed to spin.

## 7. Landé's contribution to Pauli's discovery of the Exclusion Principle

The 1925 discoveries of the exclusion principle and of electron spin have been reviewed by numerous authoritative sources, among them by the primary actors themselves[19,80]. Herein, we focus on Landé's well-acknowledged key contributions to these discoveries.

We begin by noting that Bohr thought about the shell structure of atoms and the significance of closed shells as early as 1913, guided by the periodic properties of the chemical elements. In 1920/21, he presented his findings in the Wolfskehl Lectures at Göttingen[81]. For instance, his symmetry considerations and energy estimates led him to assign the following shell structure, $nk^y$ (in terms of Bohr's original, principal quantum number n, the azimuthal quantum number k, and y the number of electrons) to Argon: $1s^2\ 2s^4\ 2p^4\ 3s^4\ 3p^4$.

Following a gentle nudge from his teacher Sommerfeld, Heisenberg presented, in his very first paper[16], his own interpretation of half-integer quantum numbers: in the process of the *Aufbau* [build-up] of an atom, Heisenberg conjectured, an electron that is being added to the existing atomic core imparts ½ ℏ of its orbital angular momentum kℏ to the core and keeps an angular momentum (k−½)ℏ. In 1924, Heisenberg joined forces with Landé and analyzed the term structure of the higher-order multiplets[82]. They concluded that if a ground-state ion with angular momentum J captures an electron, two kinds of fragments are formed, one with R=J−1/2 and another with R=J+1/2, and called this phenomenon *Verzweigung* [bifurcation].

The half-integer angular momentum R of the core, consistent with Landé's earlier views presented above, was further scrutinized by Pauli, who was bitten by the bug of the AZE while with Bohr at Copenhagen during 2022-23. In his seminal paper[69], received on 2 December 1924, Pauli set out to find out whether the magnetic moment of the core was due to a relativistic effect. Concentrating on the K-shell (characterized by Bohr's principal quantum number n=1) that was regarded as a prime suspect for both core's angular momentum and anomalous magnetism, Pauli investigated the dependence of the gyromagnetic ratio of the K-shell electrons on nuclear charge, Z, for choice atoms with well-established Zeeman spectra (alkalis, barium, mercury, thallium). He found a relativistic correction $[1-\alpha^2 Z^2/(2n^2)]$, with $\alpha$ the fine structure constant, that evaluated to as much as 18% for heavy atoms such as mercury or thallium. Since Landé's g-factor – that depends only on the angular momentum quantum numbers and not on Z – renders the Zeeman spectra with a 1% accuracy, as noted. Pauli resolved the apparent contradiction by drawing the "most natural conclusion," namely that the angular momentum and hence any magnetic contribution of the K-shell – or any other closed shell – is zero. Thus, the angular momentum R attributed to the core by Landé – and needed in the g-factor – had to be reassigned to the valence (luminous) electron(s). In Pauli's words[69], the AZE is due

> *to a peculiar not classically describable two-valuedness* [Zweideutigkeit] *of the quantum theoretical properties of the valence electron.*

Landé's work was Pauli's lodestar throughout. On 30 June 1924, Pauli wrote to Landé, Ref.[17]:

> *I am very enthusiastic about your new work, I congratulate you on it and admire your courage to speak out so boldly, although you know exactly what madness it is. I am very satisfied that the discrepancies between empiricism and previous principles of quantum theory (which already came to light with the anomalous Zeeman effect) are initially getting worse. This is often the case in physics before the real solution comes up.*

On 6 December 1924, four days after his first "two-valuedness" paper[69] was received, Pauli reported to Sommerfeld in a letter from 6 December 1924 about the progress he made on the question of the closure of electron shells and noted that Sommerfeld's emphasis in the fourth edition of *Atombau und Spektrallinien*[83] on the work of Edmund Clifton Stoner (1899-1968) was particularly helpful[84]:

> *The model ideas are now in a serious, fundamental crisis, which I believe will eventually end with a further radical sharpening of the contrast between classical and quantum theory. As can be seen from Millikan's and Landé's findings on the representability of the optical alkali doublets using relativistic formulas, the idea of definite, unambiguous orbits of the electrons in the atom is unlikely to be sustainable. With all models, one now has the strong impression that we are speaking a language that is not sufficiently adequate for the simplicity and beauty of the quantum world…. In connection with my considerations on the influence of relativity corrections on the Zeeman effect, of which I wrote you briefly last time (I have since sent them to the ZS. f. Phys.), I tried, also in the theory of the complex structure of the optical spectra to take seriously the abolition of the angular momentum of the noble gas shells. Although I am far from being able to explain all the fundamental difficulties associated with the complex structure and the anomalous Zeeman effect, I have made further progress on a number of points. This relates in particular to the question of the closure of the electron groups. Your book helped me a lot in this regard, namely by emphasizing the work of E. C. Stoner in the foreword of your book. I had completely overlooked this work when I got my hands on this Phil. Mag. But no sooner had I read your preface* *than I ran into the library and read Stoner's work. I agree with the correctness of the modification of Bohr's scheme proposed herein I am convinced and I fully agree with you that this work represents great progress. Stoner's proposal now fitted extremely well with my considerations mentioned above. I think I can justify this physically better (by considering the statistical weights of the stationary states). I also have a specific suggestion for a generalization of Stoner's approaches so that not only the electron counts in the closed shells, but also the number of realization possibilities (j-values) of open shells with given numbers of electrons can be given. In particular, the following is a simple interpretation of the omission of the triplet s term with the same principal quantum number as the normal state (singlet S term) for the alkaline earths. As a generalization of this, in the higher columns of the natural system of the elements, certain multiplier terms fall away or coincide with the smallest principal quantum number (in which electrons are already present.) In the simplest cases my theory is correct (e.g., similarity of the seven bowls to alkali like them appears in the X-ray spectra).*

In his paper[85] from October 1924 on the classification (quantum number assignment) of electron states, Stoner noted:

*This classification has been put forward by Landé. In contradiction to the older schemes, such as that of Sommerfeld, it gives a satisfactory selection principle (k changes by 1, J by 1 or 0) and at the same time brings out clearly the analogy between x-ray and optical spectra. ... Landé, in two recent papers has traced out the analogy in quantitative detail. He shows that the relativity and optical doublet separations can be both represented by the same general formula, and places beyond doubt that the two types of doublet are essentially similar in origin. ... In the classification adopted the remarkable feature emerges that the number of electrons in each completed level is equal to the double the sum of the inner quantum numbers* [known as Stoner's rule] *as assigned, there being in the K, L, M, N levels with 2, 8 (2 + 2 + 4), 18 (2 + 2 + 4 + 4 + 6), 32 electrons ... The present scheme, then, accounts well for the chemical properties; it differs from Bohr's in the final distribution suggested and in the fact that inner sub-groups are completed ... In the atoms there is only one electron external to a core composed of a completed system of electronic groups... Electrons can enter a sub-level until all orbits are occupied. ...* For there is then one electron in each possible equally probable state [our emphasis].

Pauli then took it from there in his second "two-valuedness" paper[86], in which he ascribed *Zweideutigkeit* to every electron in an atom which he characterized by an additional, fourth quantum number (apart from n, k, and m). Pauli's summary amounts to a formulation of the *exclusion principle*:

*There can never be two or more equivalent electrons in an atom for which in strong fields the values of all* [four] *quantum numbers coincide* [are the same]. *If there is an electron in the atom whose* [all four] *quantum numbers (in the external field) have certain values, then this state is "occupied."*

The reference to strong fields has to do with the Paschen-Back limit, in which the AZE takes a simpler form, cf. Eq. (8), relied on in Pauli's considerations. The introduction of the two-valued fourth quantum number then immediately explained Stoner's rule.

Sommerfeld reacted in a letter to Pauli on 18 June 1925, Ref.[84]:

*Dear Pauli! These days I have used your work on the shell completion (for my special lecture) and I am very impressed by it. You got a lot further with it than I did. From your principle of the 4 quantum numbers and from your experience with the alkali Z [eeman] effects, you can prove the Stoner classification with j = 0 and determine the existing or missing terms. That is very nice and undoubtedly correct.*

In his Nobel lecture *Exclusion Principle and Quantum Mechanics*[19] presented in 1946, Pauli described his work on the exclusion principle as follows:

*On Bohr's invitation, I went to Copenhagen in the autumn of 1922, where I made a serious effort to explain the so-called «anomalous Zeeman effect», as the spectroscopists called a type of splitting of the spectral lines in a magnetic field which is different from the normal triplet. On the one hand, the anomalous type of splitting exhibited beautiful and simple laws and Landé had already succeeded to find the simpler splitting of the spectroscopic terms from the observed splitting of the lines. The most fundamental of his results thereby was the use of half-integers as magnetic quantum numbers for the doublet-spectra of the alkali metals. On the other hand, the anomalous splitting was hardly understandable from the standpoint of the mechanical model of the atom, since very general assumptions concerning the electron, using classical theory as well as quantum theory, always led to the same triplet.*

*A closer investigation of this problem left me with the feeling that it was even more unapproachable. We know now that at that time one was confronted with two logically different difficulties simultaneously. One was the absence of a general key to translate a given mechanical model into quantum theory which one tried in vain by using classical mechanics to describe the stationary quantum states themselves. The second difficulty was our ignorance concerning the proper classical model itself which could be suited to derive at all an anomalous splitting of spectral lines emitted by an atom in an external magnetic field. It is therefore not surprising that I could not find a satisfactory solution of the problem at that time. I succeeded, however, in generalizing Landé's term analysis for very strong magnetic fields, a case which, as a result of the magneto-optic transformation (Paschen-Back effect), is in many respects simpler. This early work was of decisive importance for the finding of the exclusion principle. ...*

*On the basis of my earlier results on the classification of spectral terms in a strong magnetic field the general formulation of the exclusion principle became clear to me. The fundamental idea can be stated in the following way: The complicated numbers of electrons in closed subgroups are reduced to the simple number one if the division of the groups by giving the values of the four quantum numbers of an electron is carried so far that every degeneracy is removed. An entirely non-degenerate energy level is already «closed», if it is occupied by a single electron; states in contradiction with this*

*postulate have to be excluded. The exposition of this general formulation of the exclusion principle was made in Hamburg in the spring of 1925, after I was able to verify some additional conclusions concerning the anomalous Zeeman effect of more complicated atoms during a visit to Tübingen* [Landé] *with the help of the spectroscopic material assembled there.*

At the time of his discovery of the exclusion principle, Pauli was rather averse to any concrete physical interpretation of the two-valued fourth quantum number[81]. Enter George Uhlenbeck (1900-1988). According to his testimonial[87]:

*… it occurred to me that since (as I had learned) each quantum number must mean that the electron had an additional degree of freedom – in other words the electron must be rotating!*

Henceforth, the fourth quantum number has been known as the projection quantum number $m_S$ of the electron spin $S=1/2$. At the same time, Landé's g-factor of the atom core (g=2) was replaced by the gyromagnetic ratio of the electron, $g_S$, cf. Eq. (9) and the concomitant text. The remaining conceptual puzzle involving the orbital velocity of the spinning electron was resolved by Llewellyn Thomas (1903-1992) during his stay with Bohr in Copenhagen[88].

Thereafter, even Pauli accepted the spinning electron[81]. Pauli's winding path to the spin was captured in a letter, Figure 9, from Thomas to Samuel Goudsmit (1902-1978) from 25 March 1926, written four months after Goudsmit and Uhlenbeck submitted their take on electron spin as internal angular momentum of the electron[89]:

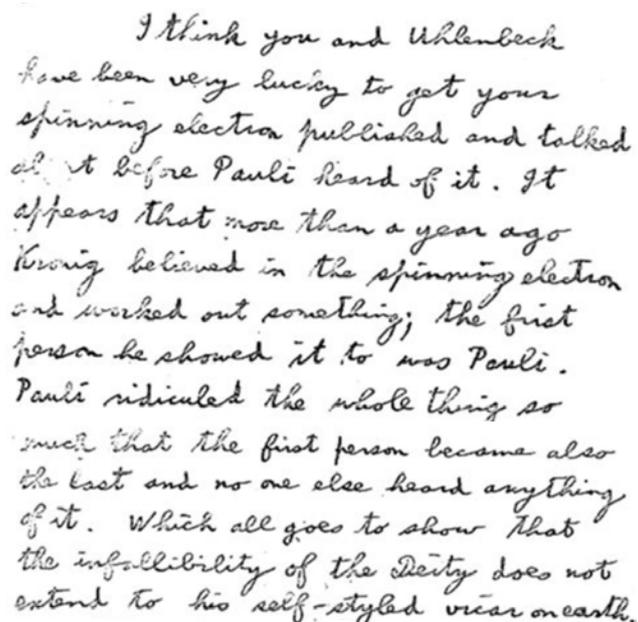

Figure 9: Excerpt from a letter written by Llewellyn Thomas to Samuel Goudsmit on 25 March 1926, Ref.[80]. Folder 6 Box 2, Samuel A. Goudsmit Papers. Niels Bohr Library & Archives, American Institute of Physics. One Physics Ellipse, College Park, MD 20740
https://repository.aip.org/islandora/object/nbla%3A16527#page/38/mode/1upOur.

We note that the g factor of the electron serves today to accurately determine the fine-structure constant[90].

## 8. Conclusions

The Zeeman effect, in particular its wide-spread variety, termed nevertheless *anomalous*, confused and confounded the pioneers of atomic physics. In Pauli's words[91], "how can one look happy when … thinking about the anomalous Zeeman effect". Upon his entry into the fray, in 1919, Alfred Landé gradually modified the sets of quantum numbers introduced by Bohr, Sommerfeld, and Debye to characterize atomic states. By adapting the concept of vector addition of angular momenta to the case of quantized electronic angular momenta of atoms, Landé came up with an organizing principle that made it possible to capture both the patterns and the subtleties of atomic Zeeman spectra amassed on the eve of the discovery of electron spin in 1925. This organizing principle was based on the g-factor, whose preliminary form Landé introduced in 1921 and kept refining until 1923. In the process, Landé attributed half-integer values to the quantum number R characterizing the angular momentum of the atomic core. The accuracy of Landé's g-factor (wihin 1%) served as a reliable guide to Pauli on his path to reassigning Landé's half-integer quantum number of the core to the outer electron -- and henceforth to the exclusion principle.

Landé's g-factor lives on as an emblem of atomic physics encapsulating the coupling of atomic angular momenta.

**Appendix A: 1962 Interview of Landé by Thomas Kuhn and John Heilbron**

In 1962, Landé was interviewed by Thomas Kuhn (1922-1996) and John Heilbron (*1934). Landé remembered the unriddling of the AZE as follows[92]:

Landé: *And from that to the vector model is only a small step. This is already a vector model -- two axes precessing around their common resultant… But I think that this paper of mine here, "Eine Quantenregel fur die räumliche Orientierung von Elektronenringen" may be the first in which this model is used extensively …*

*Well, the angular momentum always played the leading role in quantization, in Sommerfeld's and Wilson's quantum rule. This is much more important than the quantization of energy…*

*And one tried this and that, and it gradually became clearer that these quantum numbers could be associated with a vector model… Some people think more in models, and other people more in terms of mathematical symmetries, matrices. My case is only to think in models, certainly. I am not a mathematician.*

Heilbron: *Do you recall by any chance what kind of model you were thinking of which helped get the g factor.*

Landé: *Oh yes, the g factor quite at the end of this whole vector business… The only model consideration in the case of the g-factor was that there was something -- the core -- which had twice as much magnetic moment than it ought to have. Of course there were model considerations, the whole vector model is a model… This is here the first paper on the anomalous Zeeman effect, 1921* [Ref.[11]]. *… All of them depended only on having the* *magnificent photograph of Back, who analyzed each term exactly -- "This is 3/4 and this is 5/6 and this is 7/8" -- with the greatest accuracy. This is very simple. I came to Tübingen… in October 1922. Back gave me his material, already evaluated, and two months later I had the g formula -- in December 1922. This was very simple. ...*

Kuhn: *Doesn't the Sommerfeld 1920 paper already start talking combination principles?*
Landé: *You see, two things were in my head, as I remember. First of all the vector model, which of course pertains only to the spectral energies and spectral terms; on the other hand, I had in my head the anomalous Zeeman types with their intrinsic regularities. But in some way these two things had to be combined, and it was just a flash. But the strange thing is that Sommerfeld, who also had exactly the same material, and had already written the paper about the normal Zeeman effect, from spectral terms, did not get this idea first. I think one of the reasons was that as always older people are driven to think in certain fixed lines, once and for all. They cannot get away from them. And I was rather ignorant of what could be done and what should be done …*

*I had studied very thoroughly the book by Konen, Spectrum Lines [Ref.[93]], which gave all the material. Also you find all the deviations from the (Balmer) formula, and the whole material including the rules of the -- was the Zeeman effect also included I am not sure. Anyway, all this is found in Konen's book …*

Kuhn: *Didn't you after this paper find some aspects of his use of the Rumpf [core] useful in your own vector model?*

Landé: *No. I had the vector model, in which there is the orbital angular momentum in one direction, the core momentum in another, which is already two. This first paper in 1921 on the g factor -- the core momentum apparently showed anomaly… From the model, as this 1923 paper shows, if the R has double magnetic moment then the g factor should have been $J^2 + R^2 - K^2 / 2J^2$. I remember that several times I had discussions with Back, and said, that he must have made a mistake in his Zeeman types. "They ought to contain not the fraction 5/6 but the fraction 7/8." And Back refused this absolutely, and said, "You must be wrong, my figures are certainly correct." And it took quite a struggle between me as a theorist and Back as an experimentalist to convince me that there must be some modification. And then finally I drew up this table here with my expected figures from the model -- I had this model with the double magnetic moment already -- with my expected pictures; and on another sheet I took Back's figures. And then I compared Back's with mine, and then I saw the correction is $J^2$ minus 1/4. In this way I arrived at the correct g factor, which of course can be written in a more symmetric way, J times J plus 1, and so on. You see theoretical physics is just -- in this case -- all kinds of number mysteries, until finally you put some system in them. .....*
Kuhn: *When people talk about electron spin -- this was true very shortly afterwards -- everybody says: Abraham [Ref.[94]] had, shown that a rotating charge would have a gyro-magnetic ratio twice that of the normal Larmor precession.*

Landé: *I never knew about it.*

**Appendix B: A brief biography of Landé** (adapted from Ref.[14])

Alfred Landé was born on 13 December 1888 in Elberfeld (today a part of the city of Wuppertal) into a liberal Jewish family. His father Hugo Landé (1859-1936) and mother Thekla, neé Landé (1864-1932) were first cousins, Figure 10. The father was the floor leader of the Sozialdemokratische Partei Deutschlands (SPD) in Elberfeld. He was involved in drafting SPD's "Erfurt Program" aimed at improving workers' lives rather than at precipitating a socialist revolution. The mother became in 1919 one of the first female members of parliament in Rhineland. Some of Alfred Landé's ancestors served as rabbis; several are buried at the Old Jewish Cemetery in Prague. Alfred was the eldest of four siblings and was considered a prodigy in mathematics and physics. He graduated from a humanistic high school in Elberfeld at Easter 1908. By that time, Alfred became also an accomplished pianist; later on, he would earn a living for a while as a piano teacher. In 1908, he entered university to study physics and mathematics (1st semester in Marburg, 2nd-4th semester in Munich, and 5th-8th semester in Göttingen). In January 1912, he passed a state examination in Göttingen, whereby he qualified to teach physics, mathematics, and chemistry at high school. In 1912, he joined Arnold Sommerfeld (1868-1951) in Munich as a PhD student in theoretical physics with the dissertation topic "On the Method of Natural Oscillations in Quantum Theory." After two semesters, he became, on Sommerfeld's recommendation, a special assistant to David Hilbert (1862-1943) in Göttingen – with the assignment to keep him [Hilbert] abreast of the developments in physics. In parallel, he completed his doctoral thesis under Sommerfeld and received his Ph.D. from Munich in July 1914, just weeks before the outbreak of World War One. Whereupon he was drafted to serve with the Red Cross on the Eastern Front and subsequently transferred to Berlin – through the mediation of Max Born (1882-1970) whom he knew from Göttingen – to the Artillery Testing Commission (A.P.K.), which was run by Rudolf Ladenburg (1882-1952) and Max Born. Still during the war, he investigated jointly with Born the compressibility of crystals, that led them to the conclusion that atoms have volume. In December 1918, Landé took the job of a music teacher at the Odenwald School in Heppenheim while continuing his work in theoretical physics.

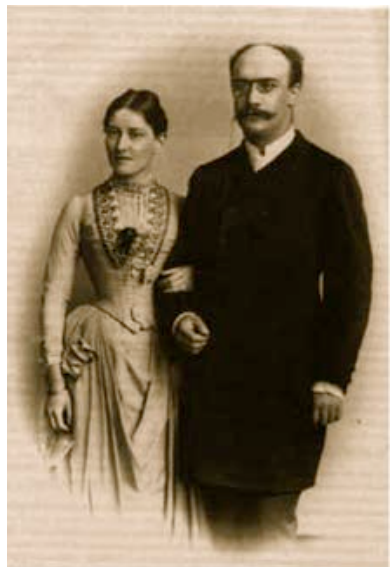

Figure 10: Alfred Landé's parents, Thekla and Hugo Landé. Elberfeld, about 1890. Wikipedia, Creative Commons.

After Max Born succeeded Max von Laue (1879-1960) at the University of Frankfurt in 1919, he hired Landé as his assistant, alongside with Otto Stern (1888-1969) and Elisabeth Bormann (1895-1986), Figures 11 and 12. The same year, Landé completed his habilitation thesis "Quantum Theory of the Helium Spectrum" and was appointed Privatdozent on October 28, 1919. On September 17, 1920, he received the *Venia Legendi* in Frankfurt. Since 1919, Landé was preoccupied with the structure of atoms and from 1920 on with the Zeeman effect. During his time in Frankfurt, he discovered what we call today Landé's g-factor. In 1922, Landé married Elisabeth Grunewald, with whom he had two sons, Arnold and Carl. In October 1922, he accepted a call to become an Extraordinarius at Tübingen. In 1929, Landé was invited to lecture at the Ohio State University (OSU) in Columbus. After a repeated stay at OSU in 1931, he accepted a professorship there, Figure 13. He remained at OSU until his retirement on 1 October 1959. Figure 14 shows Lande at the age of about 57. Landé published over 150 papers dealing almost exclusively with quantum physics isssues, as well as 10 books and 4 handbook articles. Since about 1950, he was engaged in debates on the interpretation of quantum mechanics. Landé's two sisters, Charlotte (1890-1977) and Eva (1901-1977), thanks to his help, were able to emigrate to the U.S. before the outbreak of World War Two. However, his brother Franz (1893-1942) stayed put and was murdered in Auschwitz. Their father committed suicide in 1936 after escaping from the Nazis to Switzerland. Landé died in Columbus on 30 October 1976.

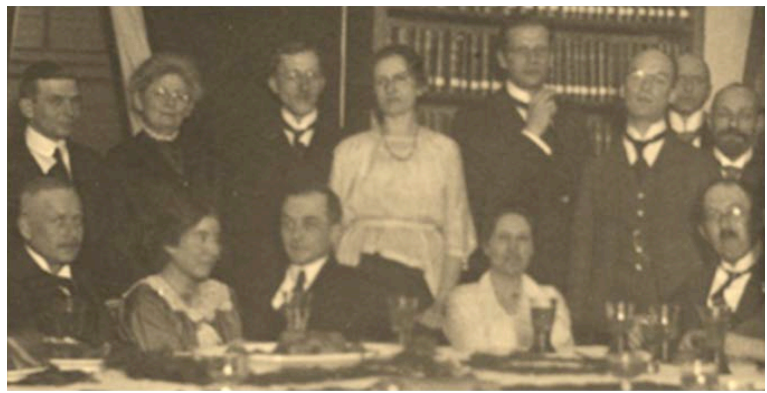

Figure 11: Frankfurt's physics faculty in 1920. Sitting from right: Otto Stern, Max Born and Richard Wachsmuth (1868-1941); standing 3rd and 4th from the right Alfred Landé and Walther Gerlach. Courtesy of the Otto Stern Estate.

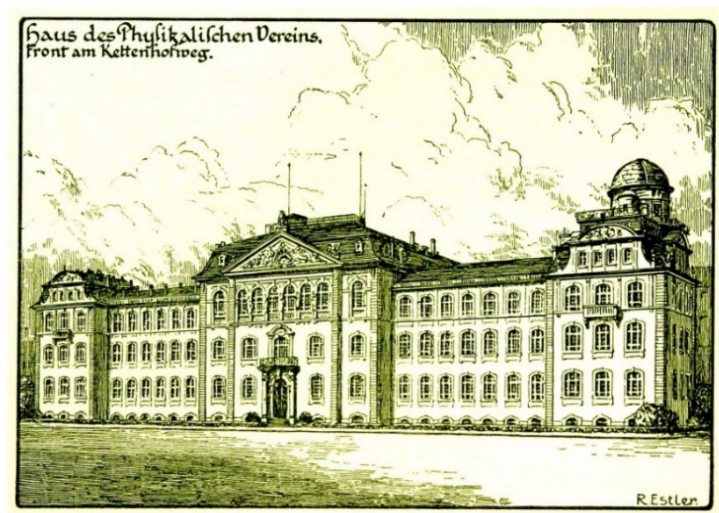

Figure 12: Historic Physics Institute of the University Frankfurt, Robert-Meyer-Str. 2 in Frankfurt (about 1920). Max von Laue, Max Born, Erwin Madelung (1881-1972) used, consecutively, the room on the second floor, 3rd window from the right as their office (Born shared his office with Landé; the Stern-Gerlach-experiment was performed in the room on the second floor, 2nd window from the right. Courtesy Universitätsarchiv Frankfurt am Main.

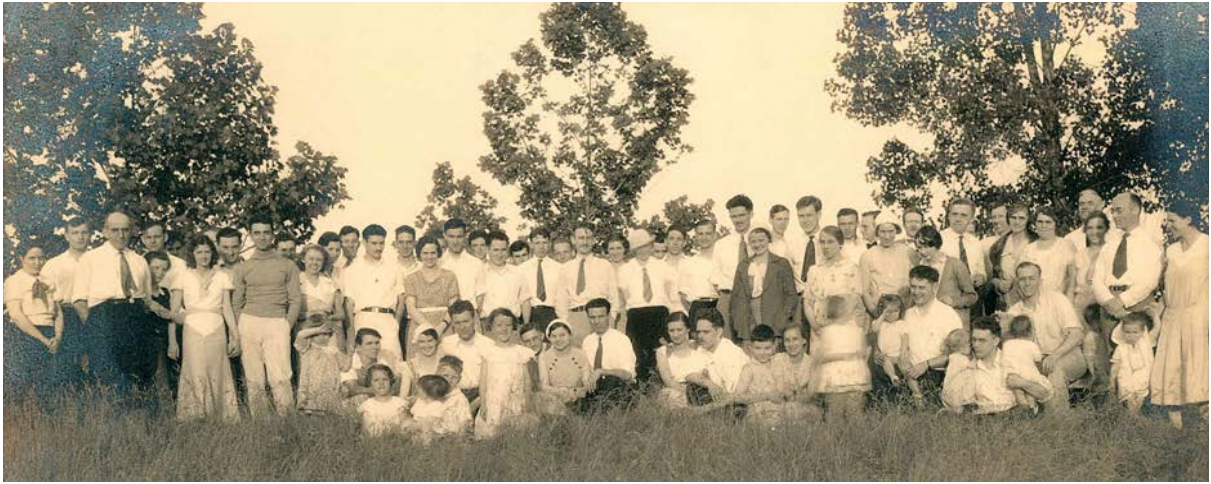

Figure 13: The physics faculty of the Ohio State University with their family members (ca. 1930). Alfred Landé and his family are on the far-right. Private communication from Louis DiMauro, The Ohio state University. Courtesy The Ohio State University.

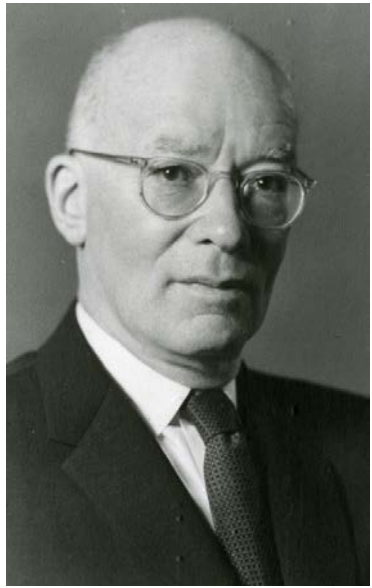

Figure 14: Alfred Landé ca. 1945. Private communication by Louis DiMauro, The Ohio State University. Courtesy The Ohio State University.

## Acknowledgments

We thank Gernot Gruber (Frankfurt), Gerard Meijer (Berlin), Dan Kleppner (Cambridge, MA), and Dudley Herschbach (Cambridge, MA) for helpful discussions. B.F. gratefully acknowledges the hospitality of John Doyle and Hossein Sadeghpour during his stay at Harvard

Physics and at the Harvard & Smithsonian Institute for Theoretical Atomic, Molecular, and Optical Physics (ITAMP).